\documentclass[%
 reprint,
 amsmath,amssymb,
 aps,
prb,
nofootinbib
]{revtex4-2}

\usepackage{graphicx}
\usepackage{dcolumn}
\usepackage{bm}
\usepackage{amssymb} 
\usepackage{multirow}
\usepackage{amsmath}
\usepackage{mathtools}
\usepackage[dvipsnames,x11names]{xcolor}
\usepackage[pdfusetitle]{hyperref}

\usepackage{nicematrix}
\NiceMatrixOptions{cell-space-top-limit = 2pt, cell-space-bottom-limit = 1.5pt}

\definecolor{Red}{rgb}{0.9, 0, 0}
\definecolor{Blue}{rgb}{0, 0, 0.9}
\definecolor{Rosa}{RGB}{224, 174, 224}

\usepackage{etoolbox}
\apptocmd{\sloppy}{\hbadness 3000\relax}{}{} 

\hypersetup{
  colorlinks,
  citecolor=blue,
  linkcolor=blue,
  urlcolor=black}

\makeatletter
\newcommand{\fmarki}{*}
\newcommand{\fmarkii}{\ensuremath{\dagger}}
                
\def\@fnsymbol#1{{\ifcase#1\or \fmarki\or \fmarkii \else\@ctrerr\fi}}
\makeatother
\renewcommand{\fmarki}{\ensuremath{\dagger}}
\renewcommand{\fmarkii}{\ensuremath{\dagger\dagger}}

\def\Vhrulefill{\leavevmode\leaders\hrule height 0.7ex depth \dimexpr0.4pt-0.7ex\hfill\kern0pt}

\newcommand{\bra}[1]{\left<\vphantom{\sum}#1\right|}
\newcommand{\ket}[1]{\left|\vphantom{\sum}#1\right>}

\newcommand{\brB}[1]{\hspace{-1.5 pt}\left(#1\right)}

\newcommand{\I}{\mathrm{i}}

\newcommand{\muB}{\mu_\mathrm{B}}

\renewcommand{\vec}[1]{\mathbf{#1}}

\begin{document}

\title{Giant orbital magnetization in two-dimensional materials} 

\author{Martin Ovesen}
 \email{martov@dtu.dk}
\author{Thomas Olsen}
 \email{tolsen@fysik.dtu.dk}
 \thanks{corresponding author}
\affiliation{Computational Atomic-scale Materials Design, Department of Physics, Technical University of Denmark, DK-2800 Kongens Lyngby, Denmark}

\date{\today}

\begin{abstract}
\noindent Orbital magnetization typically plays a minor role in compounds where the magnetic properties are governed by transition metal elements. However, in some cases, the orbital magnetization may be fully unquenched, which can have dramatic consequences for magnetic anisotropy and various magnetic response properties. In the present work, we start by summarizing how unquenched orbital moments arise from particular combinations of crystal field splitting and orbital filling. We exemplify this for the cases of two-dimensional (2D) VI$_3$ and FePS$_3$, and show that Hubbard corrections as well as self-consistent spin-orbit coupling are crucial ingredients for predicting correct orbital moments from first principles calculations. We then search the Computational 2D Materials Database (C2DB) for monolayers having tetrahedral or octahedral crystal field splitting of transition metal $d$-states and orbital occupancy that is expected to lead to large orbital moments. We identify 112 monolayers with octahedral crystal field splitting and 62 monolayers with tetrahedral crystal field splitting and for materials with partially-filled $t_{2g}$ bands, we verify that inclusion of Hubbard corrections as well as self-consistent spin-orbit coupling typically increases the magnitude of predicted orbital moments by an order of magnitude.
\end{abstract}

\maketitle

\section{Introduction}
\noindent
Magnetism at the atomic level can be understood through contributions from spin and orbital magnetic moments. Spin magnetization, like electric charge, is an intrinsic property of the electron and can be described in terms of a local magnetization density $\mathbf{m}(\mathbf{r})$. This is a well-defined field that characterizes the many-body ground state and may, for example, be calculated in the framework of density functional theory (DFT) \cite{von1972local}. Moreover, the response of the spin magnetization under a dynamic external field is governed by the fundamental (spin) magnetic excitations, which can be calculated from either many-body perturbation theory \cite{muller_acoustic_2016, costa_nonreciprocal_2020, olsen_unified_2021} or time-dependent DFT as has been applied extensively on simple bulk magnets such as Fe, Co and Ni \cite{pajda_ab_2001,buczek_different_2011,rousseau_efficient_2012,cao_ab_2018,singh_adiabatic_2019,skovhus_dynamic_2021} or more complicated crystals such as bulk NiO and MnO \cite{binci_magnons_2025} and monolayer Fe$_3$GeTe$_2$ \cite{skovhus_minority_2024}. The magnetization density is typically localized in the vicinity of atoms and the dynamics may often be accurately approximated by a simple site-based model with inter-site interactions calculated from first principles. Such models are routinely used to predict magnon energies as well as critical temperatures for magnetic order and provide a rather direct link to inelastic neutron scattering experiments, which are typically interpreted in terms of site-based models.

On the other hand, orbital moments are not intrinsic to the electron but are instead a direct consequence of the motion of electrons. The angular momentum operator $\hat{\vec{L}}$ involves the position operator $\hat{\vec{r}}$ which makes the orbital moments significantly more complicated to calculate in periodic systems, since $\hat{\vec{r}}$ is not itself periodic and is thus an ill-defined operator. This problem has largely been solved with the development of the modern theory of orbital magnetization\cite{xiao2005berry,thonhauser2005orbital,ceresoli2006orbital,aryasetiawan2019modern}, which renders the bulk orbital magnetization well-defined, despite the subtleties associated with the position operator. The theory has also been generalized to a local formulation, making it possible to calculate unit cell resolved orbital moments \cite{bianco_orbital_2013}. The modern theory thus provides a rigorous route to calculate the orbital magnetization as a unit cell average, but unlike the spin magnetization it is not possible to define a genuine continuum field for the orbital magnetization, and it is not always straightforward to assign orbital moments to the individual atoms in the unit cell. Alternatively, orbital moments are often well described by an atom-centered approximation where the Bloch states are expanded in atom-centered orbitals, on which it is straightforward to evaluate expectation values of the orbital angular momentum. This approach is expected to work as long as the orbital magnetization mainly originates from atom-centered loop currents. Finally, in addition to the computational subtleties, the orbital moments are usually quite small, more than an order of magnitude smaller than the spin moment \cite{meyer1961experimental}. For these two reasons, the orbital contribution to the total magnetization is often ignored and is typically regarded as a small correction without significant qualitative consequences for the magnetic properties.

In 2016 it was demonstrated that monolayer FePS$_3$ exhibits magnetic order below 118 K \cite{lee2016ising}. The Néel temperature is almost the same as in bulk (van der Waals bonded) FePS$_3$ implying that interlayer magnetic interactions play a minor role in stabilizing magnetic order. This is somewhat surprising, since the Mermin-Wagner theorem states that continuous symmetries cannot be broken spontaneously in 2D \cite{mermin1966absence} and magnetic order in the monolayer can thus only originate from magnetic anisotropy. However, magnetic anisotropy is usually much weaker than magnetic exchange interactions and one would therefore typically expect a somewhat lower Néel temperature in the monolayer limit. On the other hand, FePS$_3$ is known to have an unusually high magnetic anisotropy \cite{lancon_magnetic_2016,lancon_magnetic_2018, coak_emergent_2021,lee2023giant} - much larger than what is typically expected from compounds containing $3d$ transition metal elements. Magnon energies obtained from inelastic neutron scattering, for example, show that the anisotropy is an order of magnitude larger than for the isostructural compound NiPS$_3$ \cite{lancon_magnetic_2018}. It is also worth comparing with the ferromagnet CrI$_3$, which like FePS$_3$ crystallizes in a honeycomb lattice of magnetic atoms. This compound has gained attention as being the first example of ferromagnetic order in the monolayer limit \cite{huang2017layer} and the material was carefully chosen for such demonstration due to the large spin-orbit coupling (and associated magnetic anisotropy) induced by the heavy I atoms. Yet, the anisotropy is almost an order of magnitude smaller \cite{cenker_direct_2021} than the case of FePS$_3$, which does not contain any heavy elements.

In Ref. \cite{olsen2021magnetic} the magnetic properties of monolayer FePS$_3$, NiP$_3$, and MnPS$_3$ were calculated from first principles using the DFT+U method with non-self-consistent spin-orbit interactions. Although most of the results were found to agree well with experimental properties of the bulk compounds, the magnetic anisotropy of FePS$_3$ was found to be an order of magnitude lower than predicted by experiments. Shortly after this, a similar study of the same compounds found that performing DFT+U calculations with {\it self-consistent} spin-orbit coupling yielded a fully unquenched orbital magnetization and an associated large magnetic anisotropy in FePS$_3$. The orbital moment is pinned to the atomic plane and the spin-orbit coupling $\sim \mathbf{L}\cdot\mathbf{S}$ thus induces a strong uniaxial anisotropy. The crucial point here is that the large unquenched orbital moment can only be obtained from self-consistent SOC. Even if SOC is rather weak, the effect of self-consistency yields an order of magnitude difference for the magnetic anisotropy. Due to the crucial importance of magnetic anisotropy in 2D materials it is pertinent to include such effects correctly in first principles calculations. That is, to include spin-orbit coupling self-consistently whenever this is required.

In the present work, we start by explaining the basic mechanisms that lead to fully unquenched orbital moments in compounds containing transition metal elements. It arises as a combination of crystal field splitting, SOC and (sometimes) the Mott-Hubbard mechanism where strong correlation opens a gap in a metallic band. We focus on tetrahedral and octahedral field splitting of the $d$-bands. We then select all such monolayers from the Computational 2D Materials Database (C2DB) and demonstrate significant differences (often an order of magnitude) between orbital moments calculated from self-consistent and non-self-consistent approaches.\vfill

\section{Theory}\label{sec:theory}
\subsection{Crystal field splitting and the filling of orbitals}
\begin{figure}
    \centering
    \includegraphics[width=1\linewidth]{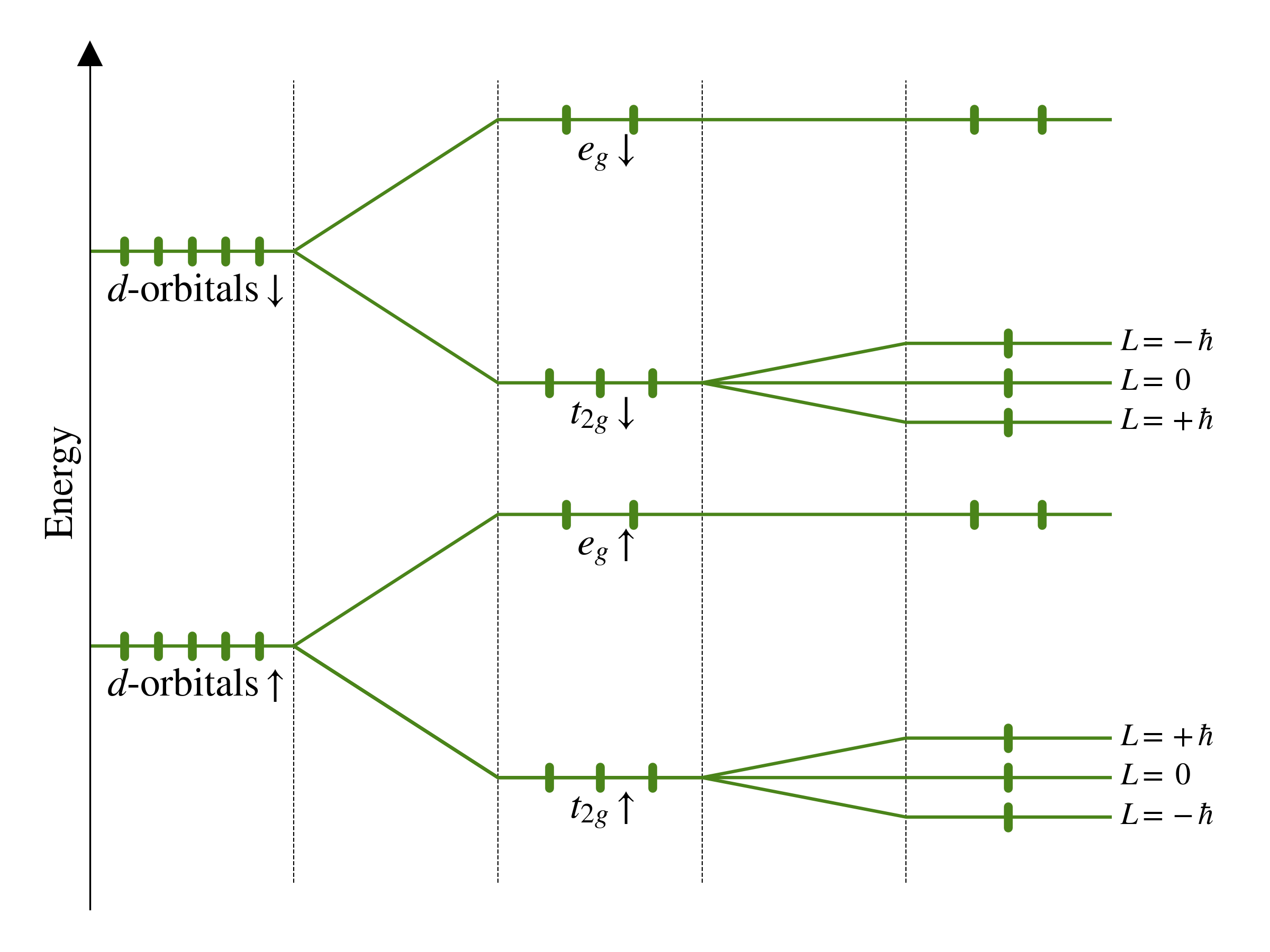}
    \caption{Sketch of the first-order splitting of the $d$-orbital energy levels in an octahedral crystal field (middle) and with the spin--orbit interaction included (to the right).}
    \label{fig:splitting_atom}
\end{figure}
\noindent For the isolated transition metal atom, the $d$-orbitals form a 10-fold degenerate energy level if relativistic effects are neglected. 
In solids the degeneracy is lifted by the crystal field and the resulting energy splitting may be approximated by idealized high symmetry configurations of the environment \cite{bernhardt2025introduction,companion1964crystal}. For example, transition metals are commonly bonded to either 4 or 6 equidistant ligands, which result in tetrahedral and octahedral crystal fields respectively. This leads to splitting of the energies into triple degenerate (for one spin channel) $t_{2g}$ and doubly degenerate $e_g$ levels. If the system is magnetically ordered the single-particle energy levels (at the mean-field level) will be subject to an exchange splitting between the two spin channels. Moreover, spin--orbit coupling may give rise to additional splitting within the crystal field energy levels and favor occupancy of bands with a finite orbital angular momentum. This is illustrated in Fig. \ref{fig:splitting_atom}.
\begin{figure*}
\centering
\begin{minipage}{0.3\textwidth}
\raggedright\textbf{(a)}
\includegraphics[width=1.\linewidth]{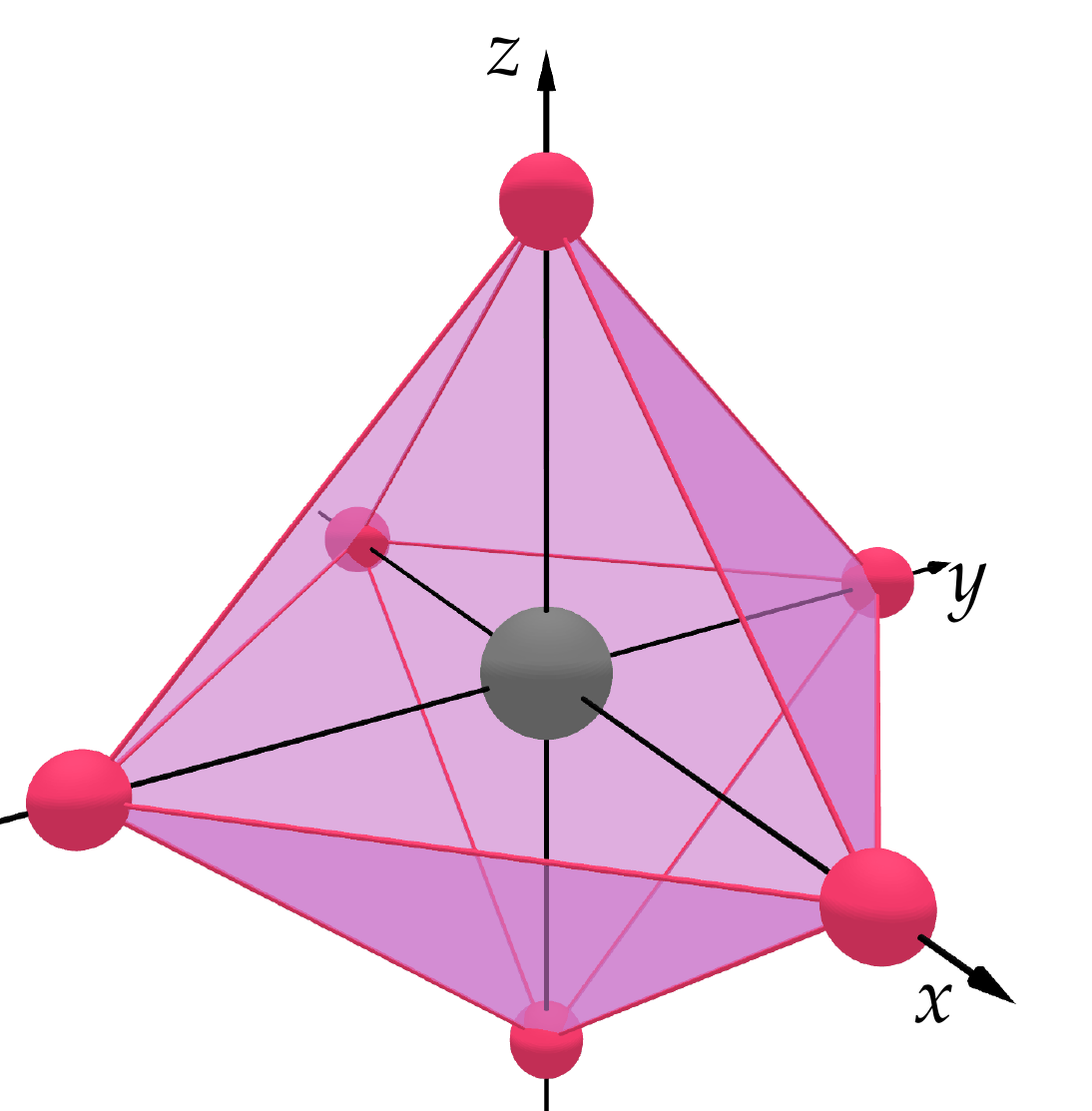}
\end{minipage}\hfill
\begin{minipage}{0.3\textwidth}
\raggedright\textbf{(b)}
\includegraphics[width=1.\linewidth]{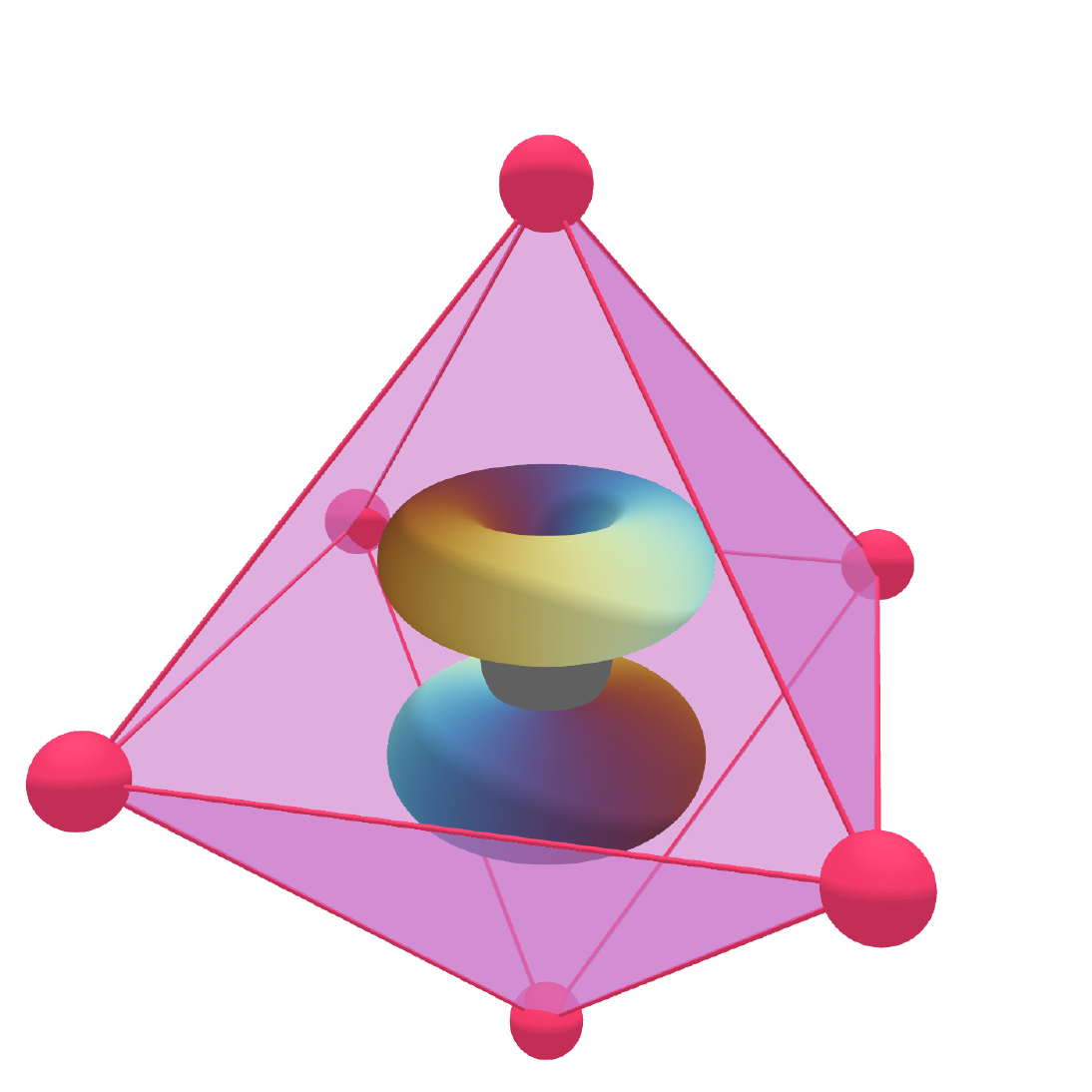}
\end{minipage}\hfill
\begin{minipage}{0.3\textwidth}
\raggedright\textbf{(c)}
\includegraphics[width=1.\linewidth]{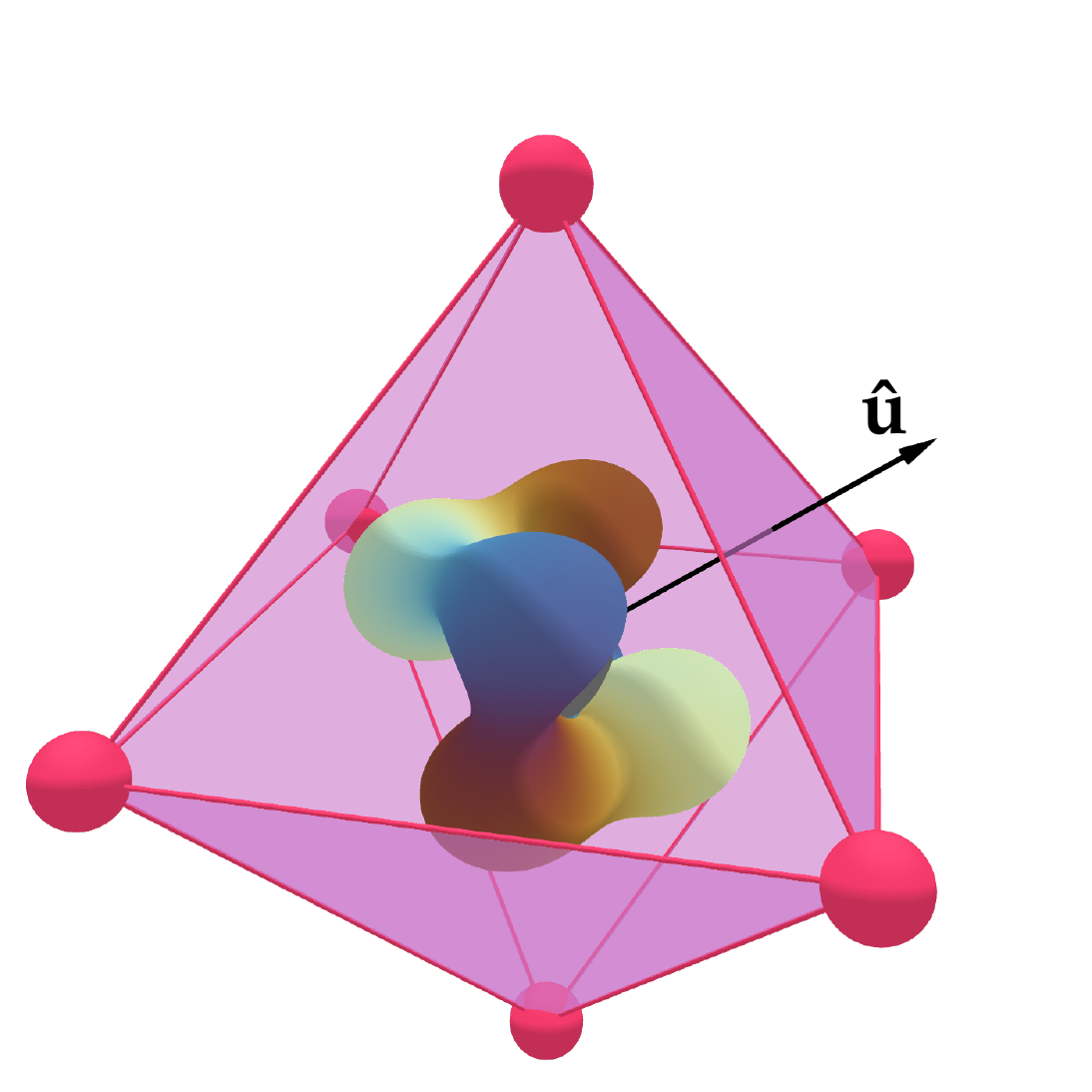}
\end{minipage}
\caption{\textbf{(a)} Geometry and coordinate system of a transition metal atom (grey) inside the octahedral cage consisting of six ligand atoms (pink). \textbf{(b)}-\textbf{(c)} Respectively, the states $\ket{\psi_{\hat{\vec{z}}}}$ and $\ket{\psi_{\hat{\vec{u}}}}$ in the octahedral cage. The cyclic color gradient displays the winding of the wave function's complex phase which is necessary for nonzero orbital moments.}
\label{fig:octahedral_cage}
\end{figure*}

In the following we focus on the perfect octahedral crystal field model. This is represented by considering the transition metal atom at the origin and the six ligands placed at
\begin{equation*}
    \brB{\pm a,\ 0,\ 0},\qquad\brB{0,\ \pm a,\ 0},\qquad\mathrm{and}\qquad\brB{0,\ 0,\ \pm a}
\end{equation*}
for some interatomic distance $a$, see Fig. \ref{fig:octahedral_cage}. In this coordinate system, $t_{2g}$ states are spanned by the $\left\{d_{xy},\ d_{yz},\ d_{xz}\right\}$ orbitals and $e_g$ states by the $\left\{d_{x^2-y^2},\ d_{z^2}\right\}$ orbitals. In the basis of real spherical harmonics, the diagonal matrix elements the angular momentum operator vanishes. In addition,
\begin{equation}
    \bra{d_{x^2-y^2}}\hat{L}_\alpha\ket{d_{z^2}}=0\qquad\mathrm{for}\qquad\alpha=x,y,z
\end{equation}
which implies that the states in the $e_g$ level cannot carry orbital moments and therefore cannot be split by the spin--orbit interaction. On the other hand, it is straightforward to pick linear combinations of the $t_{2g}$ states that carry finite angular momentum. For example,
\begin{equation}\label{eq:psi_z}
    \ket{\psi_\vec{\hat{z}}}=\frac{-1}{\sqrt{2}}\brB{\ket{d_{xz}}+\I \ket{d_{yz}}}=\ket{Y_{2}^1}
\end{equation}
is the familiar eigenstate of the $\hat{L}_z$ operator, illustrated in Fig. \ref{fig:octahedral_cage}, with an eigenvalue of $+\hbar$. Similarly, one may construct states that have eigenvalues of $\pm \hbar$ along the $x$ or $y$ direction. It is thus possible to form eigenstates of angular momentum pointing towards any of the vertices of the octahedral cage. However, for most 2D materials having octahedral crystal field splitting (for example FePS$_3$), the atomic plane is parallel to a face of the octahedron and populating a single eigenstate of orbital angular momentum (pointing towards a vertex) would thus break the symmetry of the monolayer. On the other hand, having orbital angular momentum normal to a octahedral face (and thus perpendicular to the atomic plane) conserves the in-plane rotational symmetries and will typically be favored. In particular for the case of FePS$_3$, DFT calculations have demonstrated a large out-of-plane orbital angular momentum, which is perpendicular to an octahedral face and cannot be explained by the state \eqref{eq:psi_z}. In terms of the coordinate system shown in Fig. \ref{fig:octahedral_cage} the direction of angular momentum is given by $\vec{\hat{u}}=\brB{\vec{\hat{x}}+\vec{\hat{y}}+\vec{\hat{z}}}/\sqrt{3}$. One might suppose that the orbital moment in FePS$_3$ can be understood as a simple rotation of $\psi_\vec{\hat{z}}$ such that the orbital moment then points along $\vec{\hat{u}}$. The rotation operator implementing this transformation is
\begin{equation}
\hat{R}_{\vec{\hat{z}}\to\vec{\hat{u}}}=\exp\brB{-\I\vartheta\vec{\hat{n}}\cdot\vec{\hat{L}}/\hbar}
\end{equation}
with an axis of rotation $\vec{\hat{n}}=\brB{\vec{\hat{y}}-\vec{\hat{x}}}/\sqrt{2}$ and an angle of rotation $\vartheta=\arccos\brB{1/\sqrt{3}}\approx54.74^\circ$. Writing $\hat{R}_{\vec{\hat{z}}\to\vec{\hat{u}}}$ in the basis
$\left\{\begin{matrix} d_{xy}, & d_{yz}, & d_{z^2}, & d_{xz}, & d_{x^2-y^2}\end{matrix}\right\}$ yields
\begin{equation*}
\hat{R}_{\vec{\hat{z}}\to\vec{\hat{u}}}=\frac{1}{3}
\begin{bmatrix}
    \phantom{-}2 & \phantom{-}1 & \sqrt{3} & \phantom{-}1 & \phantom{-}0 \\[4 pt]
    -1 & \phantom{-}\dfrac{\sqrt{3}-1}{2} & \sqrt{3} & -\dfrac{\sqrt{3}+1}{2} & \phantom{-}\sqrt{3} \\[8 pt]
    \phantom{-}\sqrt{3} & -\sqrt{3} & \phantom{-}0 & -\sqrt{3} & \phantom{-}0 \\[8 pt]
    -1 & -\dfrac{\sqrt{3}+1}{2} & \sqrt{3} & \phantom{-}\dfrac{\sqrt{3}-1}{2} & -\sqrt{3} \\[8 pt]
    \phantom{-}0 & -\sqrt{3} & \phantom{-}0 & \phantom{-}\sqrt{3} & \phantom{-}\sqrt{3}
\end{bmatrix}.
\end{equation*}
Clearly the rotated state $\hat{R}_{\vec{\hat{z}}\to\vec{\hat{u}}}\psi_\vec{\hat{z}}$ will not be spanned by only the $t_{2g}$ basis functions and therefore cannot be used to explain the orbital moment in FePS$_3$. Instead consider the state
\begin{equation}\label{eq:psi_u}
\begin{split}
    \ket{\psi_\vec{\hat{u}}}&=
    \frac{1}{\sqrt{3}}\ket{d_{xy}} \\
    &\hspace{19 pt}-\brB{\frac{1}{2\sqrt{3}}+\frac{\I}{2}}\ket{d_{yz}}-\brB{\frac{1}{2\sqrt{3}}-\frac{\I}{2}}\ket{d_{xz}}
\end{split}
\end{equation}
also visualized in Fig. \ref{fig:octahedral_cage}. This is clearly a $t_{2g}$ state, and although it is not an eigenstate of any component of the angular momentum operator, it does have the required expectation value
\begin{equation}\bra{\psi_\vec{\hat{u}}}\vec{\hat{L}}\cdot\vec{\hat{u}}\ket{\psi_\vec{\hat{u}}}=\hbar.
\end{equation}
The physical differences between eigenstates of the angular momentum operator and states like $\psi_\vec{\hat{u}}$ can be further highlighted by considering their respective loop-currents as represented by the winding of the complex phase. As Fig. \ref{fig:octahedral_cage} shows, eigenstates of an angular momentum operator such as $\phi_\vec{\hat{z}}$ of the $\hat{L}_z$ operator represent looping currents that follow perfectly circular paths which lie in planes perpendicular to the $z$-axis whereas the looping current of $\psi_\vec{\hat{u}}$ must follow a more undulating path, maximizing the distance between the electron and the repulsive electron clouds near the ligands as the $t_{2g}$ states do by construction.

In 2D materials spin-orbit coupling may thus make it advantageous to form certain linear combinations of high angular momentum (parallel or perpendicular to the atomic plane) if the $t_{2g}$ band is partially filled. In solids, the single-particle orbitals are never eigenstates of the angular momentum operator and the distinction between states such as Eq. \eqref{eq:psi_z} and Eq. \eqref{eq:psi_u} becomes irrelevant - only expectation values of the orbital angular momentum are well defined. In this regard Eq. \eqref{eq:psi_u} should simply be regarded as a qualitative explanation of how fully unquenched orbital angular momentum may be obtained along the vertices as well as faces of the octahedra. Finally, we remark that the notation $d_{xy}$, $d_{xz}$, and $d_{yz}$ for the states $t_{2g}$ entails the risk of slight confusion for 2D materials. This is because the Cartesian basis is always with respect to the octahedral vertices, whereas one typically defines the $z$-axis as being orthogonal to the atomic layers and thus orthogonal to the faces of the octahedra.

\subsection{Hubbard corrections in the framework of noncollinear-spin DFT}
\noindent Commonly used exchange-correlation functionals such as the local spin density approximation (LSDA) \cite{von1972local} or the Perdew-Burke-Ernzerhof (PBE) \cite{perdew1996generalized} functional suffer from delocalization errors and are often incapable of describing correlated materials accurately. 
This deficiency is intimately related to the total energy not being piecewise linear as a function of orbital occupation \cite{cohen_challenges_2012}. A simple and widely used method for correcting this is the addition of an atom-centered Hubbard-like penalty $E_U$ to the total energy functional, which cancels the quadratic dependence of the occupation \cite{anisimov_density-functional_1993,liechtenstein_density-functional_1995}. The rotationally invariant implementation suggested by Dudarev et al. \cite{dudarev1998electron, dudarev2019parametrization} defines the penalty functional as
\begin{equation}\label{eq:hubbard}
    E_U=\frac{U}{2}\sum_a\mathrm{tr}\brB{\rho_a-\rho_a\rho_a}
\end{equation}
where $U$ is a tunable positive parameter controlling the strength of the penalty and $\rho_a$ is the atomic single-particle density matrix (in both spin and orbitals) which characterizes the occupancy of the orbitals involved in the correction. $\rho_a$ is Hermitian and positive semi-definite, and generally $\rho_a$ is expected to satisfy $\mathrm{tr}\brB{\rho_a}\geq\mathrm{tr}\brB{\rho_a\rho_a}$, which ensures that $\mathrm{tr}\brB{\rho_a-\rho_a\rho_a}\geq0$ and therefore also $E_U\geq0$. In the case where $\rho_a$ is an idempotent matrix, i.e. $\rho_a\rho_a=\rho_a$, the Hubbard penalty clearly vanishes. An idempotent density matrix represents a system of states with no fractional occupations (eigenvalues of idempotent matrices are exclusively $0$ or $1$) and the method thus introduces missing correlation effects by energetically penalizing fractional occupation of atomic orbitals and pushing $\rho_a$ toward idempotency. In particular, the Hubbard correction leads to increased localization of orbitals and may open a gap in a partially filled band. We note that it is possible to calculate the Hubbard parameter $U$ self-consistently \cite{cococcioni_linear_2005}, but here we will mainly be interested in the qualitative effects that give rise to unquenched orbital magnetism and we will choose $U$ somewhat arbitrarily in the calculations below. 

As will be shown below, the Hubbard correction \eqref{eq:hubbard} is typically required to (correctly) open a gap in a partially filled $t_{2g}$ band. For magnetic systems, spin-orbit coupling (SOC) will then favor orbital angular momentum that is anti-aligned the spin of the occupied band. Since SOC is a rather weak effect, it often suffices to add it non-selfconsistently as a post processing step \cite{heide_describing_2009, olsen2016designing}. However, such a procedure is not capable of yielding the correct orbital polarization \cite{kim2021magnetic, ovesen2024orbital}. The reason is that an initial DFT+U calculation will largely yield the correct band structure with a gapped $t_{2g}$ band, but the occupied part of the band will largely be an arbitrary linear combination of $t_{2g}$ states and there is no way SOC can yield a polarized state non-self-consistently. In order to obtain the correct orbital polarization in the ground state one has to apply non-collinear calculations with self-consistent SOC.

\begin{figure}
    \centering
    \includegraphics[width=1\linewidth]{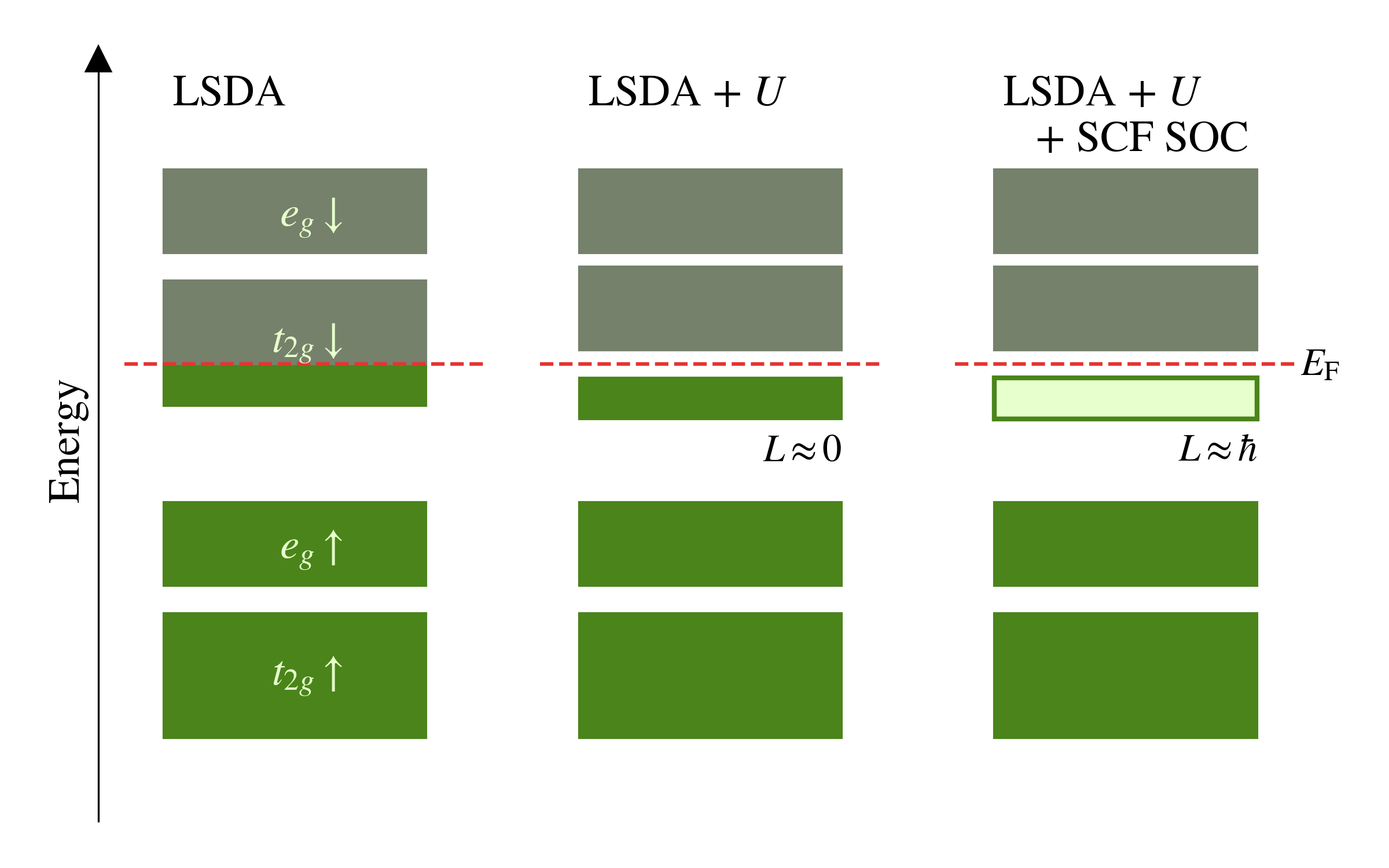}
    \caption{Sketch of the band structure of a material like FePS$_3$ with an octahedral crystal field and six valence electrons occupying the $d$-shell. Pure LSDA yields a metallic ground state whereas LSDA+$U$ opens a gap, but with no orbital magnetization. Only the LSDA+$U$ with self-consistent SOC yields a ground state with the correct orbital magnetization.}
    \label{fig:splitting_crystal}
\end{figure}

\begin{figure*}[t]
    \centering
    \includegraphics[width=0.7\linewidth]{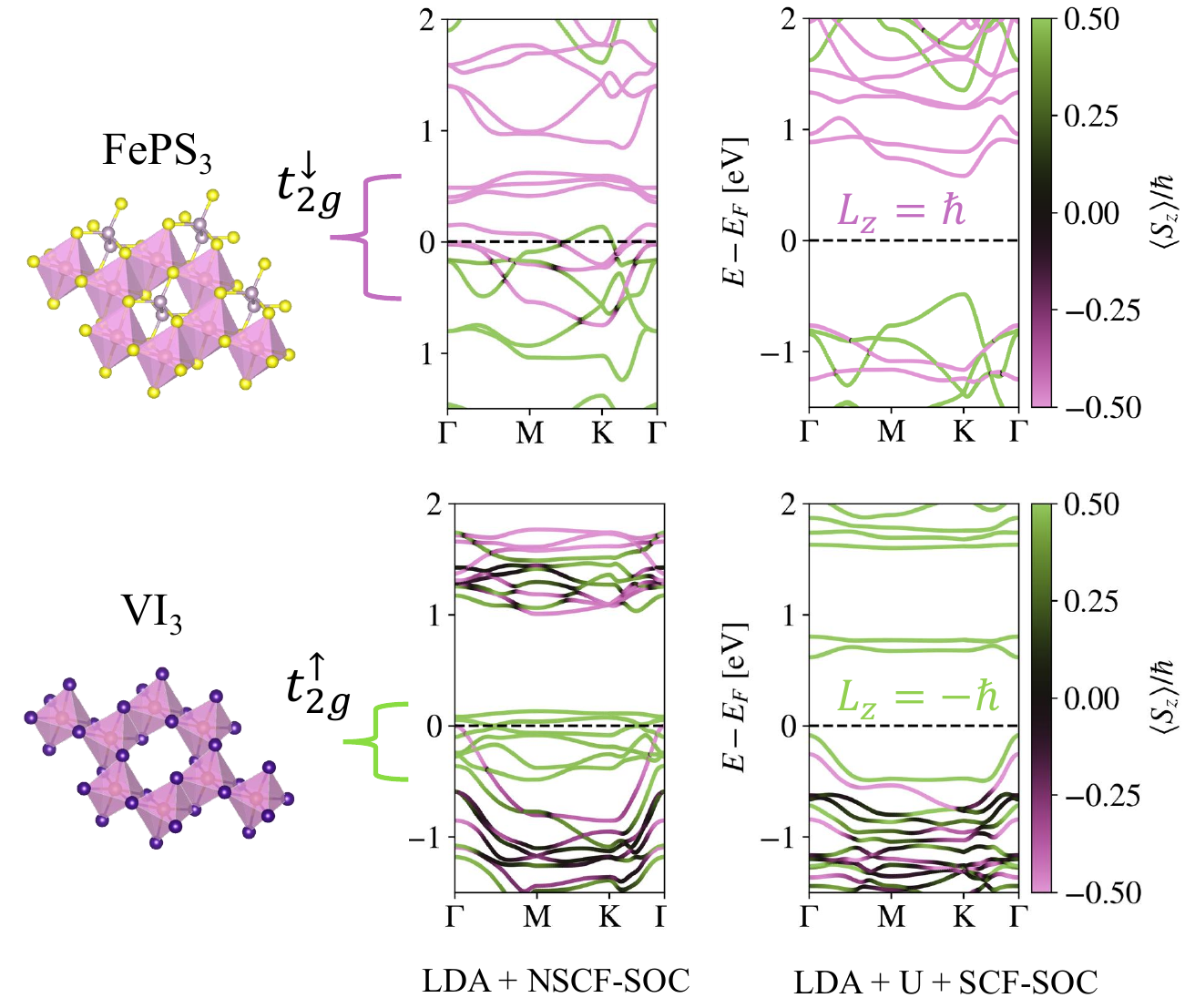}
    \caption{Band structures of FePS$_3$ (top) and VI$_3$ (bottom) with and without self-consistent SOC and Hubbard corrections. In both cases, the partially occupied $t_{2g}$ bands are indicated in the bare LDA calculations. Note that there are two magnetic atoms pr unit cell and the $t_{2g}$ bands thus have six states for {\it each} spin channel.}
    \label{fig:FePS3,VI3}
\end{figure*}


Previous works \cite{kim2021magnetic,hovancik2023large,ovesen2024orbital} have discussed the layered van der Waals compounds VI$_3$ and FePS$_3$; both measured to have unusually large atomic orbital moments of approximately 1 $\muB$ in magnitude. Due to the octahedral crystal field surrounding the transition metals and number of valence electrons in the $d$-shell, these orbital moments can be understood as originating from partially-filled $t_{2g}$ bands. However, these materials become metallic with small orbital moments when simulated with spin-polarized DFT with the simple local spin density and PBE exchange-correlation functionals. Adding Hubbard corrections to the transition metals opens the band gap, but their orbital moments remain small. Finally, including both a Hubbard correction and self-consistent SOC yields a finite gap \text{and} an unquenched orbital moment. In Fig. \ref{fig:splitting_crystal} the mechanism is illustrated schematically for the case of FePS$_3$ which has six valence electrons in the $d$-shell of the Fe atom.



\section{Computational details}\label{sec:computational}
\noindent In general we predict that any tetrahedrally or octahedrally coordinated material will exhibit large (almost fully unquenched) orbital magnetization if the $t_{2g}$ band is partially occupied and if Hubbard corrections are able to open a gap in the band. Moreover, it is expected that inclusion of self-consistent SOC will be required to predict the correct ground state angular momentum. In order to validate these predictions we have selected a large set of materials 
from the Computational 2D Materials Database (C2DB) \cite{haastrup2018computational,gjerding2021recent}. Octahedral field splitting occurs rather commonly, but for simplicity we have restricted ourselves to two prototypes: The first kind consists of materials with three atoms in the unit cell and with the symmetry of layer group 72 (p$\overline{3}$m1), i.e. 1T-phase transition metal dichalcogenides or dihalides such as FeI$_2$ or MnTe$_2$. The second kind are transition metal trihalides with the symmetry of layer group 71 (p$\overline{3}$1m), i.e. compounds such as such as CrI$_3$ or VI$_3$ where the transition metal atoms are arranged in a honeycomb lattice. All selected materials thus have a central transition metal bonded to six ligands that form an octahedral cage. We note that this selection excludes well known 2D magnets such as FePS$_3$ (used as a prototypical example in the present work) and CrGeTe$_3$ \cite{gong_discovery_2017}, which both have layer group p$\overline{3}$1m as well. The reason for this is that the counting of residual valence electrons for the transition metal atom becomes much simpler for pure halides or chalcogenides and the analysis in terms of partially filled $t_{2g}$ bands is easier for such compounds. In addition, we note that all transition metal tetrel trichalcogenides (such as CrGeTe$_3$) in the C2DB involve Cr and have a completely filled majority $t_{2g}$ band and thus no significant orbital magnetization. In addition, all transition metal phosphurus trichalcogenides in C2DB involve either Mn or Ni (filled $t_{2g}$ band and no orbital magnetization) or Fe or Co, which have partially filled minority $t_{2g}$ bands and exhibit orbital magnetization very similar to the case of FePS$_3$ which will be discussed in detail below. Calculations were also performed on materials with tetrahedral field splitting where we have restricted ourselves to materials which have three atoms in the unit cell with the symmetry of layer group 59 (p$\overline{4}$m2).

Each material was subjected to two different ground state calculations using the open-source electronic structure package GPAW \cite{mortensen2005,enkovaara2010,mortensen2024gpaw} using the projector augmented-wave method \cite{blochl1994projector}. The first calculation was performed with spin-polarized collinear DFT using LSDA and SOC added non-self-consistently subsequently. The second calculation used non-collinear DFT with LSDA, a Hubbard correction on the transition metal $d$-orbitals with $U=3$ eV and SOC included self-consistently. Note that the choice of $U=3$ eV is largely arbitrary as the main purpose of the Hubbard penalty is to be strong enough to split the partially-filled $t_{2g}$ states. Nevertheless, it has been demonstrated for FePS$_3$ and VI$_3$ that increasing the Hubbard penalty beyond $U=2$ eV does not alter the atomic orbital moments significantly \cite{ovesen2024orbital}. Furthermore, the initial spin configuration for each non-collinear DFT calculation was chosen as a collinear ferromagnet pointing along the previously determined easy axis of each crystal \cite{torelli2019high}. Reasonably, one might suspect that ferromagnetic and anti-ferromagnetic states may yield different magnitudes of the orbital moments; however, for the prototypical FePS$_3$ crystal this was not the case, and since giant orbital moments arise from the local ligand environment, the long-range magnetic order should not play a major role on the magnitude of the orbital moments. For both collinear and non-collinear DFT calculation, the local orbital moments were calculated with the projector augmented-wave atom-centered approximation (PAW-ACA) \cite{ovesen2024orbital}. All calculations were performed using a plane wave basis with a cutoff energy of 800 eV and a $k$-point density of at least 12 Å.

\section{Results}\label{sec:results}
\noindent Before we present a compilation of the results for the materials extracted from the C2DB, we exemplify the approach by the two prototypical compounds VI$_3$ and FePS$_3$. The former case is expected to contain 2 valence electrons on the V atom and thus a partially filled majority $t_{2g}$ band and the latter has 6 valence electrons and thus a partially filled minority $t_{2g}$ band (because the exchange splitting is larger than crystal field splitting). The pure LSDA with non-self-consistent SOC (NSCF) calculation and the LSDA+U+SOC (SCF) calculations for both compounds are shown in Fig. \ref{fig:FePS3,VI3}. In both cases the bare LSDA calculations yield metallic states as expected from partially filled bands. Performing the calculations with Hubbard correction opens a gap in both cases and thus splits the $t_{2g}$ band into an occupied part and an occupied part. If SOC is included self-consistently in the calculation, the occupied part will acquire an orbital angular momentum that is anti-aligned with the spin of that band resulting in an orbital magnetization of the order of one $\mu_\mathrm{B}$ per transition metal atom.

\begin{figure}[t]
    \centering
    \includegraphics[width=1\linewidth]{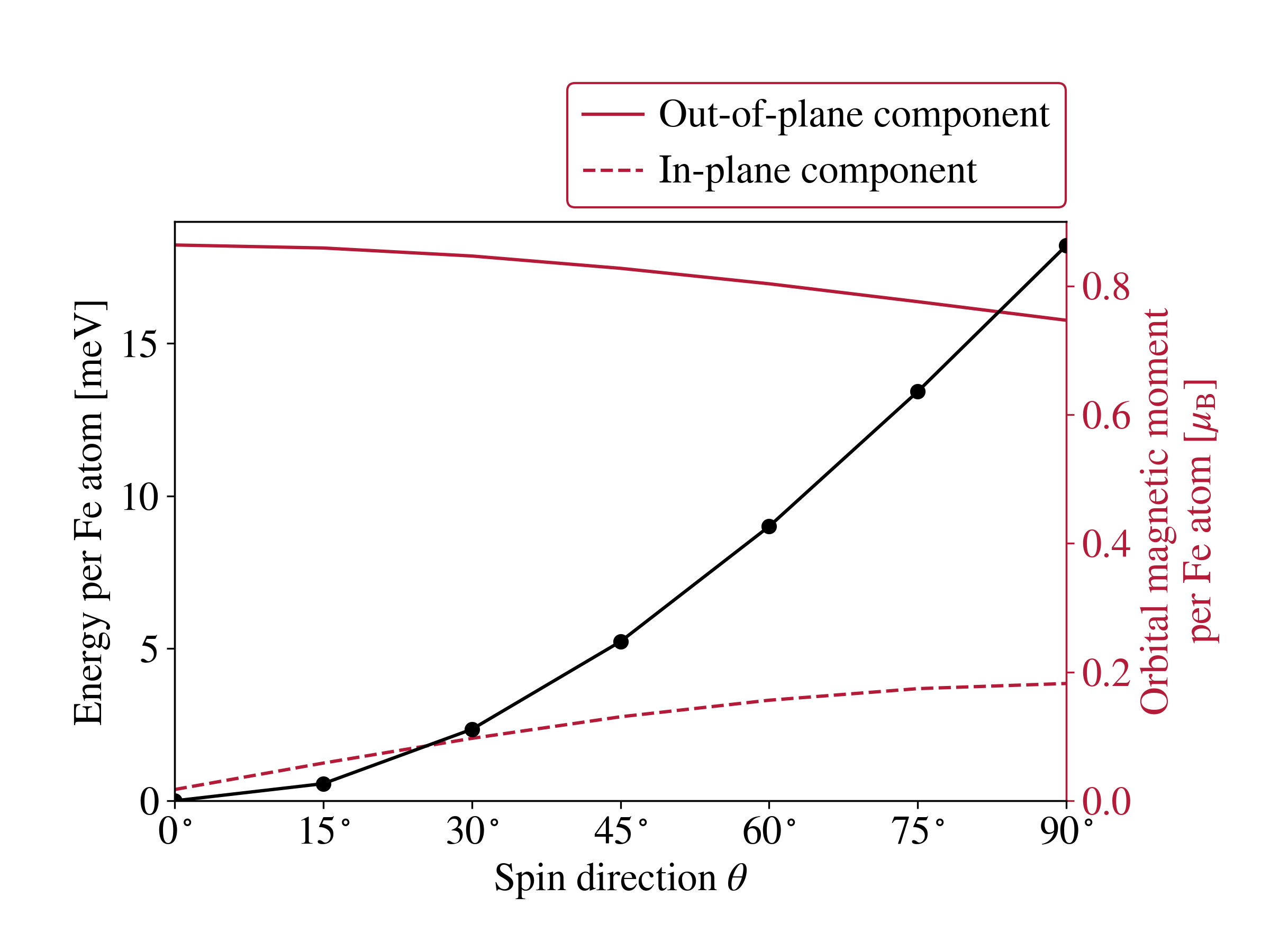}
    \caption{Energy of the FePS$_3$ crystal and components of the orbital moment vector at the Fe atoms when the direction of the spins at the Fe atoms are constrained to point along the direction $\theta$. Each data point represents a constrained-spin DFT calculation, and the energy is measured w.r.t. the $\theta=0$ data point.}
\label{fig:FePS3_CDFT}
\end{figure}

Since these unquenched orbital moments arise from the ligand geometry around a transition metal and since the SOC couple the directions of spin and orbital magnetic moments, the presence of unquenched orbital moments may signify an unusually large magnetic anisotropy as the orbital moment firmly pins the direction of spin moment to the lattice. This enhancement of the magnetic anisotropy shall be shown directly through constrained-spin DFT of the two-dimensional FePS$_3$ crystal. Through constrained-spin DFT, the direction of spins near the Fe atoms can be forced to point along directions varying between $\theta=0^\circ$ (pointing out of the crystal plane) and $\theta=90^\circ$ (pointing along the crystal plane). The resulting increase in the potential energy of the crystal as the spin direction is forced further in-plane is displayed in Fig. \ref{fig:FePS3_CDFT}. Additionally shown is also the components of the orbital magnetic moment at the Fe atoms which continue to point predominantly out-of-plane even as the direction of the local spin moments are completely in-plane. This increase in the crystal energy coupled with the mostly static direction and magnitude of the local orbital magnetic moments exemplify the connection between magnetic anisotropy and giant orbital magnetic moments.

\begin{figure*}
    \centering
    \includegraphics[width=1\textwidth]{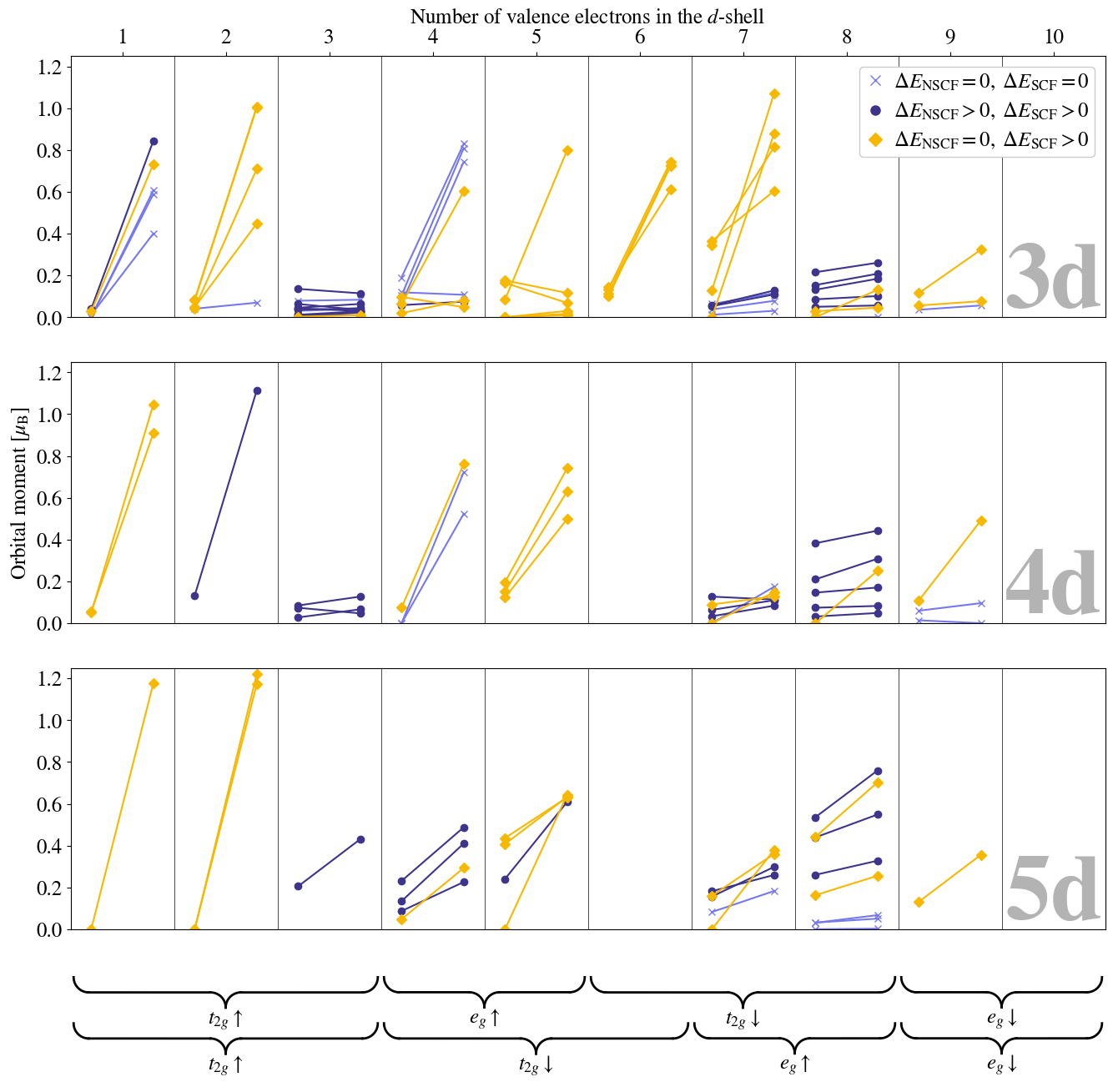}
    \caption{Increase in the orbital moment of magnetic atoms in octahedral environments  when self-consistent spin--orbit and the Hubbard correction are turned on. Each line represents a material and the left end-point indicates the NSCF calculation and the right end-point indicates the SCF calculation. The results are divided into $3d$, $4d$ and $5d$ compounds and organized in terms of the transition metal valence states. We distinguish between three situations corresponding to whether the band gap $\Delta E$ is finite or not for the NSCF and SCF calculations. No calculations resulted in $\Delta E_\mathrm{NSCF}>0$ and $\Delta E_\mathrm{SCF}=0$. Below the panels we have sketched two possible filling scenarios that may be realized: exchange splitting dominated ($t_{2g}^\uparrow\rightarrow e_{g}^\uparrow\rightarrow t_{2g}^\downarrow\rightarrow e_{g}^\downarrow$) or crystal field splitting dominated ($t_{2g}^\uparrow\rightarrow t_{2g}^\downarrow\rightarrow e_{g}^\uparrow\rightarrow e_{g}^\downarrow$).}
    \label{fig:om_increase}
\end{figure*}

In Fig. \ref{fig:om_increase} we show the results of such calculations for 112 2D materials from the C2DB having octahedral crystal field splitting. The materials include p$\overline{3}$1m trihalides as well as p$\overline{3}$m1 dichalcogenides/dihalides. The results are divided into three rows corresponding to $3d$, $4d$ and $5d$ transition metal elements and for each of these we have assigned a valency of the $d$-shell by removing one electron per halide atom and two electrons per chalcogenide atom. Each material is represented by a line with the left end-point marking the orbital moment with bare LDA (and non-self-consistent SOC) and the right end-point marking the orbital moment with self-consistent SOC and Hubbard correction. We only show the magnitudes of the orbital moment here. In the supplemental information, we include a table containing all the materials and the electronic properties (gaps, spin magnetic moments and orbital magnetic moments) for the two calculations.

\begin{figure*}
    \centering
    \includegraphics[width=1\textwidth]{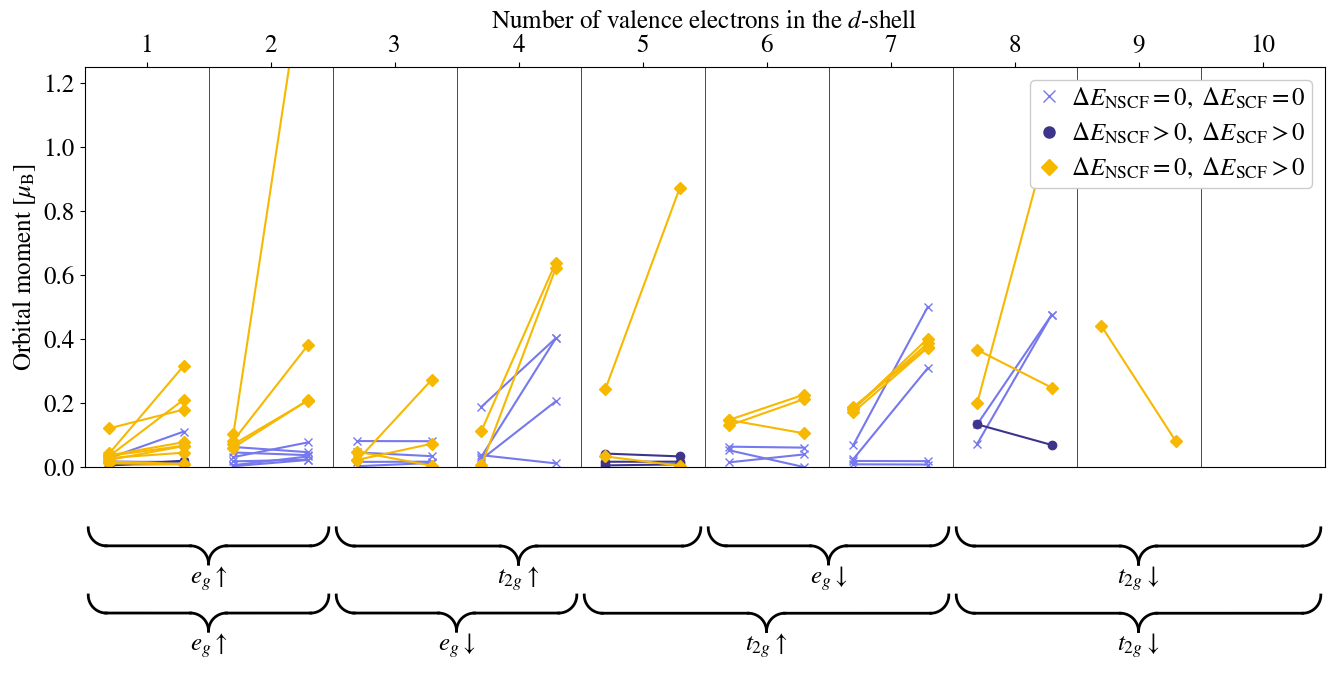}
    \caption{Increase in the orbital moment of magnetic atoms in tetrahedral environments  when self-consistent spin--orbit and the Hubbard correction are turned on. Each line represents a material and the left end-point indicates the NSCF calculation and the right end-point indicates the SCF calculation. We distinguish between three situations corresponding to whether the band gap $\Delta E$ is finite or not for the NSCF and SCF calculations. No calculations resulted in $\Delta E_\mathrm{NSCF}>0$ and $\Delta E_\mathrm{SCF}=0$. Below the panels we have sketched two possible filling scenarios that may be realized: exchange splitting dominated ($e_{g}^\uparrow\rightarrow t_{2g}^\uparrow\rightarrow e_{g}^\downarrow\rightarrow t_{2g}^\downarrow$) or crystal field splitting dominated ($e_{g}^\uparrow\rightarrow e_{g}^\downarrow\rightarrow t_{2g}^\uparrow\rightarrow t_{2g}^\downarrow$).}
\label{fig:tetrahedral}
\end{figure*}

It is straightforward to verify that the data shown in Fig. \ref{fig:om_increase} largely follow the trends discussed in Sec. \ref{sec:theory}. We may analyze the results in terms of the number of valence electrons $N_\mathrm{v}$ and starting with $N_\mathrm{v}=1,2$ we always expect a partially filled $t_{2g}$ band and a resulting large orbital moment. This is indeed what is observed: the NSCF calculations yield small (but finite due to inclusion of NSCF SOC) orbital moments and these are generally increased by an order of magnitude in the SCF calculations. For $N_\mathrm{v}=3$ one always expects a filled majority $t_{2g}$ band and the orbital moment (almost) vanishes in the SCF as well as the NSCF calculations. We also note that all of the materials with $N_\mathrm{v}=3$ are insulating in the SCF calculations as expected from the crystal field splitting, while one of the $3d$ materials are metallic in the NSCF calculations due to a small exchange splitting and overlap between majority and minority $t_{2g}$ bands. The situation becomes slightly more complicated for the materials with $N_\mathrm{v}=4,5,6,7$ because the presence or absence of a partially filled $t_{2g}$ band depends on the relative magnitudes of crystal field splitting and exchange splitting. If exchange splitting is strong, the majority $e_g$ is filled first and the orbital moments remain small. If, on the other hand, crystal field splitting is dominating, the minority $t_{2g}$ band is filled first and one obtains $S=1$ and $S=1/2$ states for $N_\mathrm{v}=4$ and $N_\mathrm{v}=5$ respectively. We see both situations in the $3d$ compounds where we clearly observe quenched, as well as fully unquenched orbital moments. The $N_\mathrm{v}=6$ compounds considered here all have partially filled minority $t_{2g}$ band since filling the minority $t_{2g}$ band before the $e_g$ band would result in a non-magnetic systems. This is the electronic structure relevant for FePS$_3$ considered above. For the $N_\mathrm{v}=7$ states we observe both quenched and unquenched orbital moments because the minority $t_{2g}$ may or may not be partially filled depending on the magnitude of exchange splitting. Similar to the $N_\mathrm{v}=4,5$ states, the electronic structure is reflected by the spin with $S=1/2$ having no orbital moments (both $t_{2g}$ bands are filled) and $S=3/2$ having fully unquenched orbital moments. The $N_\mathrm{v}=8,9$ states always have filled $t_{2g}$ bands and are therefore expected to have quenched orbital moments. We do, however, observe rather large moments in the $5d$ compounds, but these are obtained with NSCF as well as SCF and simply originate from the strong SOC in these materials rather than partially filled $t_{2g}$ bands.

In Fig. \ref{fig:tetrahedral} we have repeated these calculations on 62 2D materials having transition metal atoms with tetrahedral field splitting. Again, the $d$-states split into a $t_{2g}$ band and an $e_g$ band, but now the $e_g$ band will have the lowest energy. Thus, it is expected that unquenched moments appear for $N_\mathrm{v}=3,4,8,9$ (exchange dominated) or $N_\mathrm{v}=5,6,8,9$ (crystal field dominated). These trends are observed from Fig. \ref{fig:tetrahedral}, although less clearly compared to the case of octahedral splitting. For example, in the case of $N_\mathrm{v}=5$, most of the materials are exchange-dominated and have $S=5/2$ with filled majority bands. These do not host orbital moments. In contrast, there is a single material where both $e_g$ bands are filled (crystal field dominated) and the partially occupied $t_{2g}$ band (with $S=1/2$) acquires a large orbital moment upon inclusion of Hubbard corrections and self-consistent SOC.

We note that previous calculations of orbital moments in the C2DB carried out using the PBE functional and non-self-consistent SOC only found a single crystal in the entire C2DB having an atomic orbital moment larger than 0.5 $\muB$ \cite{ovesen2024orbital}. In contrast, by using LDA+$U$+SCF-SOC we have identified 38 crystals with octahedral field splitting and 6 crystals with tetrahedral field splitting that have local orbital moments larger than 0.5 $\muB$. 

\section{Conclusion}\label{sec:conclusion}
\noindent We have outlined a mechanism that may give rise to fully unquenched orbital moments in transition metal elements with octahedral or tetrahedral crystal field splitting. The fact that partially filled $t_{2g}$ bands yield large orbital moments is well known, and here we have focused on the particular case of 2D materials where the orbital moments tend to be orthogonal to octahedral faces rather than point towards vertices. We demonstrated that superpositions of $t_{2g}$ states can be constructed that have expectation values of $\hbar$ in a direction orthogonal to the atomic plane (and an octahedral face) although such states are not eigenstates of any angular momentum components. We exemplified this with 2D FePS$_3$ and VI$_3$, which comprise generic prototypes that exhibit partially filled minority and majority $t_{2g}$ bands, respectively, and showed that DFT correctly predicts such states {\it only} if SOC and Hubbard corrections are included self-consistently.

We then considered 112 monolayers with octahedral field splitting from the C2DB and validated that large unquenched orbital moments may be predicted from simple counting of valence electrons. Again, DFT calculations only yield the correct unquenched moments if SOC is included self-consistently. The crucial importance of self-consistent SOC has been previously reported for the case of FePS$_3$ \cite{kim2021magnetic,amirabbasi2023orbital,ovesen2024orbital}. Here we have explained that observation in detail and shown that such a requirement (self-consistent SOC) is generally expected for any material exhibiting unquenched orbital moments. 

Self-consistent SOC is generally rather demanding in terms of computational load, and it is common to add SOC non-self-consistently following a collinear SCF calculation. In particular, the band structures one may obtain with self-consistent and non-self-consistent SOC are almost identical, and it is often assumed that the effect of SOC is small and that self-consistency is irrelevant. This is definitely not true for orbital moments in materials with partially filled $t_{2g}$ bands. 

From a first principles perspective, it is rather crucial to be able to predict the correct orbital moments. If they are unquenched it may comprise 20-50 {\%} of the total magnetization and strongly influence the magnetic properties. Moreover, for 2D materials, the magnetic anisotropy is a vital ingredient for having magnetic order at finite temperatures, and the presence of unquenched orbital moments strongly influence the predicted values of magnetic anisotropy. For the case of FePS$_3$, the predicted magnetic anisotropy increases by an order of magnitude when the orbital magnetization is properly accounted for by self-consistent SOC \cite{kim2021magnetic} and it is a crucial requirement for a reliable prediction of the Néel temperature. In general, magnetic anisotropy acts as the driving force for magnetic order in 2D and scrutinizing monolayers with partially filled $t_{2g}$ band and unquenched orbital moments could therefore be a promising route for discovering new high-temperature 2D magnets.

The scope of this article has been limited to the cases of crystals with octahedral and tetrahedral field splittings; however, other ligand geometries may also be advantageous when searching for magnetic materials with large orbital moments. For example, materials with trigonal prismatic ligand geometries split the $d$-band of the transition metal into three levels, one being spanned by the $d_{xy}$ and $d_{x^2-y^2}$ states \cite{huisman1971trigonal}. A complex linear combination of these states may allow an $L_z=2\hbar$ state. We have, however, not yet managed to identify such a material in the C2DB.

\begin{acknowledgments}
\noindent The authors acknowledge funding from the Villum Foundation Grant No. 00029378. The calculations were performed on Niflheim, a high-performance supercomputer cluster at the Technical University of Denmark, which is supported by the Novo Nordisk Foundation Data Science Research Infrastructure 2022 Grant: A high-performance computing infrastructure for data-driven research on sustainable energy materials, Grant no. NNF22OC0078009.
\end{acknowledgments}

\bibliography{bibtex.bib}

\end{document}